\documentclass[twocolumn,showpacs,preprintnumbers,amsmath,amssymb,prb,floatfix]{revtex4}

\usepackage{epsfig,psfrag}
\usepackage{color}
\usepackage{graphicx}
\usepackage{dcolumn}
\usepackage{bm}
\def \rd{\mathrm{d}}

\begin{document}

\title{Signatures of Wigner molecule formation in 
interacting Dirac fermion quantum dots}

\author{Tomi Paananen$^{1}$, Reinhold Egger$^1$, and Heinz Siedentop$^2$}
\affiliation{
$^{1}$Institut f\"ur Theoretische Physik, Heinrich-Heine-Universit\"at, 
D-40225 D\"usseldorf, Germany \\
$^{2}$Mathematisches Institut, Ludwigs-Maximilians-Universit\"at M\"unchen, 
D-80333 M\"unchen, Germany}

\date{\today}
\begin{abstract}
We study $N$ interacting massless Dirac fermions confined  in 
a two-dimensional quantum dot. Physical realizations of this problem include
a graphene monolayer and the surface state of a strong topological insulator. 
We consider both a magnetic confinement and an infinite mass 
confinement.  The ground state energy is computed 
as a function of the effective interaction parameter $\alpha$ from the 
Hartree-Fock approximation and, alternatively, by employing the M\"uller 
exchange functional.  For $N=2$, we compare those approximations
to exact diagonalization results.  The Hartree-Fock energies
are highly accurate for the most relevant interaction 
range $\alpha\alt 2$, but the M\"uller functional leads to an 
unphysical instability when $\alpha\agt 0.756$.  
Up to 20 particles were studied using Hartree-Fock calculations.
Wigner molecule formation was observed for strong but realistic 
interactions, accompanied by a rich peak structure in the addition 
energy spectrum.
\end{abstract}


\maketitle

\section{Introduction}
\label{sec1}

Massless two-dimensional (2D) Dirac fermions are of central importance 
in several condensed matter applications of current
interest, in particular for monolayer graphene\cite{geim,castro}
and for the surface state of a 3D strong 
topological insulator (TI).\cite{hasan,zhang} 
These systems offer readily accessible table-top realizations of
relativistic quantum physics, where electron-electron interactions
are typically much stronger than in atomic physics. 
Interactions are characterized by an effective fine structure 
constant $\alpha$, where for graphene $\alpha \approx 1$ to $2$
depending on experimental details,\cite{castro} while 
 in TIs  $\alpha$ is probably somewhat smaller due to
the large dielectric constant of the relevant thermoelectric
materials (for instance, Bi$_2$Se$_3$ or Bi$_2$Te$_3$).\cite{thermal}
We here study the problem of $N$ massless 2D Dirac quasi-particles
confined to a circular quantum dot of radius $R$ and interacting 
through the Coulomb potential (with prefactor $\propto \alpha$).  
Quantum dots formed in 2D semiconductor heterostructures 
have been studied in much detail over the past two decades,
both experimentally\cite{kouwenhoven} and theoretically.\cite{reimann}
Given the exceptional properties of Dirac fermions and the 
unique properties of the underlying materials, it is of
considerable practical and fundamental interest to investigate 
Dirac fermion quantum dots.  Since the commonly employed 
electrostatic gating\cite{kouwenhoven} is problematic due to 
the (recently observed\cite{stander}) Klein tunneling phenomenon,
the question of how to confine Dirac fermions in a controlled manner
arises.  While quasi-bound states induced  by electrostatic
potentials have also been studied,\cite{pereira,kostya,tapash,matulis,recher}
we here consider two types of \textit{stable} confinement:
(i) an infinite-mass boundary condition\cite{berry,schnez,wurm} 
on the single-particle wavefunction at $r=R$, 
and (ii) confinement by a spatially 
inhomogeneous magnetic field profile.\cite{ademarti,peeters,hausler,jpa}
Graphene dots have already been investigated experimentally by several
groups,\cite{exp1,exp2,exp3,exp4,exp5,exp6} where confinement
was so far created lithographically. While this (approximately)
corresponds to case (i) above, such a procedure may give rise to 
uncontrolled disorder effects along the boundary, and
route (ii) may offer a promising alternative  for
future experiments, see also Ref.~\onlinecite{heinzel}.  
For the TI surface state, we are not aware of experimental reports of 
quantum dot physics, but confinement should be achievable as well using, 
e.g., suitably arranged close-by ferromagnetic layers.

On the theoretical side, another difficulty arises from the Dirac
nature of the quasi-particles when one attempts to include 
electron-electron interactions.
For the $N$-particle problem, where a first-quantized formulation generally 
offers the most natural route,\cite{reimann} the problem arises from the 
unboundedness of the single-particle Dirac Hamiltonian in Eq.~\eqref{d0} below.
This causes the Brown-Ravenhall\cite{brown} ``disease:'' 
the unbounded spectrum allows particles to lose arbitrary amounts of energy
by transferring their energy in (real) scattering events to other particles.
To circumvent this problem, suppose that the chemical
potential is located just above the Dirac point. 
We then follow Sucher\cite{sucher} and confine the Hilbert space 
to positive-energy eigenstates of the full single-particle problem,
i.e., we assume an inert filled Dirac sea. This projection approach
has been successfully employed in the same context before,\cite{hausler,jpa} 
and one can analyze also other values for the chemical potential.
The accuracy of this method was carefully assessed in Ref.~\onlinecite{hausler}. In short, the presence of a spectral gap due to confinement allows to 
implement Sucher's no-pair approximation\cite{sucher} since electron-hole pair excitations
neglected in this approach have to overcome the gap.

Below we show and compare 
results from three different computational approaches. 
In particular, we perform self-consistent Hartree-Fock (HF) calculations 
and, in addition, study a similar self-consistent 
variational procedure using the so-called 
M\"uller density matrix functional 
(replacing the Fock term).\cite{muller,siedentop} 
In atomic physics applications, the M\"uller functional is sometimes
superior to the HF approach and is valuable because
it yields a \textit{lower} bound for the 
ground-state energy.\cite{siedentop} We note that
HF calculations for graphene dots with infinite-mass 
confinement have also been carried out by other groups.\cite{hf1,hf2}
While the HF approximation is known to provide
an upper bound for the exact (within Sucher's projection approach) 
ground-state energy,\cite{jpa}  the M\"uller functional is again expected 
to generate a variational lower bound. 
We will compare results from those two approaches to exact 
diagonalization computations for $N=2$ interacting Dirac quasi-particles.  
The model and those three numerical approaches are 
described in Sec.~\ref{sec2},
while the comparison for $N=2$ can be found in Sec.~\ref{sec3}.
As expected, the exact results are always bracketed by the results obtained 
from the M\"uller functional and under the HF approximation.  
However, within the range $\alpha\le 2$
studied in this work, we find that the HF results
are much closer to the exact results and provide a rather accurate
approximation. However, results based on the M\"uller functional 
show an unphysical divergence when $\alpha\agt 0.756$, and are less
accurate than the HF results for small $\alpha$.
(Of course, for $\alpha\to 0$, all three methods recover the correct
noninteracting results.)
Having established the HF approach as highly accurate approach
for $\alpha\le 2$ and $N=2$, we continue in
Sec.~\ref{sec4} with a presentation of HF results for $N>2$ Dirac particles
in a quantum dot with infinite-mass confinement.
Besides the ground-state energy, we study various 
physical observables like the particle density or the spin density.
Our results suggest that in a confined geometry
Dirac particles can form a ``Wigner molecule''  as previously
discussed for Schr\"odinger particles in semiconductor 
dots.\cite{reimann,bedanov,egger,uzi}  When the Coulomb interactions dominate
over the kinetic energy, a Wigner crystal can be formed where 
electrons spontaneously order in a crystalline structure.  
The presence of a confining potential makes this Wigner crystallization
more favorable,\cite{egger} and although no Wigner crystal is
expected for bulk graphene,\cite{wigner} we find that the confined geometry
allows for a finite-size Wigner ``molecule'' even for Dirac fermions. 
The paper concludes with a discussion in Sec.~\ref{sec5}.

\section{Model and computational approaches}\label{sec2}

In this section we discuss the model studied in this work, 
and address the different calculational schemes employed to study
the $N$-particle problem for interacting Dirac fermions in a
 quantum dot. 

\subsection{Single-particle model}

We consider a single species of massless 2D Dirac fermions 
described by the single-particle Hamiltonian ($-e<0$ is the electron charge)
\begin{equation}\label{d0}
H_0 = v_F {\bm \sigma} \cdot \left( {\bm p}+\frac{e}{c}{\bm A} \right) 
+ M\sigma_3 ,
\end{equation}
where ${\bm \sigma}=(\sigma_1,\sigma_2)$ and the 
Pauli matrices $\sigma_i$
refer to the sublattice structure of the honeycomb lattice 
for graphene\cite{castro} 
or to the electronic spin degree of freedom 
for the TI surface state.\cite{hasan}
The Fermi velocity in graphene is $v_F\approx 10^6$~m$/$s, while
the corresponding value for the TI surface state is approximately half this
value.  A  single Dirac cone as in Eq.~\eqref{d0} can be realized for a TI 
surface,\cite{hasan} but in graphene there generally is a four-fold degeneracy
due to the valley and spin degrees of freedom.\cite{castro}  For graphene,
we then assume a spin- and valley-polarized situation where
the single-valley theory  [Eq.~\eqref{d0}] 
gives useful predictions.\cite{scatt}  
In fact, our basic qualitative conclusion, i.e., Wigner crystallization
is possible in graphene dots,  is also found from HF calculations 
including the spin and valley degrees of freedom.\cite{unpub}
In addition,
we allow for a static vector potential ${\bm A}({\bm r})$ 
corresponding to inhomogeneous magnetic fields or,
in the case of graphene, also to strain-induced pseudo-magnetic 
fields.\cite{castro}  Finally, $M({\bf r})$ corresponds to a mass
term.  In order to form a quantum dot, where Dirac fermions are confined
to a bounded spatial region, say, a disk of radius $R$ around the origin,
we now consider the two possibilities mentioned in Sec.~\ref{sec1}.
We study circularly symmetric cases, where the total angular
momentum operator $J=-i\hbar\partial_\phi+\hbar \sigma_z/2$ is conserved
and has eigenvalues $\hbar j$ with half-integer $j\equiv m+1/2$, $m\in
\mathbb{Z}$.  Single-particle solutions to $H_0 \psi=E \psi$
can then be written as
\begin{equation} \label{wave}
\psi_m(r,\phi) = e^{im\phi} \left( \begin{array}{c}\psi_{1,m}(r) \\
ie^{i\phi} \psi_{2,m}(r) \end{array}\right).
\end{equation}
In what follows, we measure energies (lengths) in units of $\hbar v_F/R$ ($R$),
and we always focus on $E>0$ solutions.

(i) \textit{Infinite mass confinement.---} 
A well-known way to describe confinement theoretically is to impose
an infinite-mass boundary condition on 
the wavefunction [Eq.~\eqref{wave}] at $r=1$, 
i.e., $M(r<1)=0$ and $M(r>1)\to \infty$.
As shown by Berry and Mondragon,\cite{berry} the effect of $M(r)$ in
Eq.~\eqref{d0} is then fully captured by the boundary condition
\begin{equation}\label{bc}
\psi_{1,m} (1) = \psi_{2,m}(1),
\end{equation}
stating that no current flows through the boundary.
With the Bessel function $J_m$, the Dirac equation is solved for $r<1$
by the \textit{Ansatz}
\[
\psi_{1,m}(r)=A J_{m}(Er),\quad \psi_{2,m}(r)=
A J_{m+1}(Er),
\]
where the boundary condition  \eqref{bc}  yields the 
energy quantization condition\cite{berry} 
\begin{equation}\label{cond}
J_m ( E_{mn} ) = J_{m+1} ( E_{mn} ).
\end{equation}
This equation has to be solved numerically. 
(Note again that $E$ is given in units of $\hbar v_F/R$ and $r$
in  units of $R$.) Positive-energy
solutions for given $m$ are then labeled by $n\in \mathbb{N}$. We mention
in passing that there are no zero-energy solutions.\cite{berry} 
The normalization factor $A$ is 
\begin{equation}\label{amn}
A_{mn} = [\pi ( J_m^2 - J_{m-1}J_{m+1} + J_{m+1}^2 - J_{m} J_{m+2})]^{-1/2},
\end{equation}
where all Bessel functions are evaluated at $E_{mn}$.   
To summarize, the single-particle solutions $\psi_a$ under a circular 
infinite-mass confinement are labeled by $a=(m,n)$ with $m\in \mathbb{Z}$
and $n\in \mathbb{N}$. The eigenenergies $E_a$ follow
 by solving Eq.~\eqref{cond} and the eigenspinor is ($r<1$)
\begin{equation}\label{confmass}
\psi_a(r,\phi) = A_a  e^{im\phi} \left( \begin{array}{c}
J_m(E_a r) \\ i e^{i\phi} J_{m+1} ( E_a r ) \end{array} \right).
\end{equation}
For a detailed discussion of the single-particle spectrum,
 see Ref.~\onlinecite{schnez}.

(ii) \textit{Magnetic confinement.---}
A second possibility to confine Dirac quasi-particles is to employ
spatially inhomogeneous magnetic fields.  This possibility has been
explored theoretically before,\cite{ademarti,peeters,hausler,jpa} and we 
study the simplest case of a piecewise constant magnetic field, 
 $B(r)=B\Theta(r-1)$ with $B>0$ and the Heaviside step function $\Theta(x)$.
The eigenenergies for this 
single-particle problem can be found numerically and were given in
Ref.~\onlinecite{jpa}.  The spectrum contains ``dot states'', with 
probability density concentrated in the disk region $r<1$, plus
relativistic bulk Landau states for $r>1$. 
The Landau states are weakly perturbed by the presence of the dimensionless 
``missing flux'' parameter $\delta := R^2/2\ell^2$,
where $\ell:= \sqrt{c/eB}$  is the magnetic length.  Because of this
perturbation, Landau level energies are slightly shifted away from their
standard bulk value, but dot states can be clearly distinguished 
in the single-particle spectrum. 
With chemical potential chosen such that all bulk
Landau states below the first Landau state, $E^{(1)} := \sqrt{2} R/\ell$ (in 
units of $\hbar v_F/R$), are filled,  
the relevant dot states are in the window $0<E_a<E^{(1)}$. 
All eigenstates can again be labeled by $a=(m,n)$, i.e., using
angular momentum $j=m+1/2$ and the index $n\in \mathbb{N}$.
For given missing flux $\delta$, there are $N_b(\delta)$ dot states, 
where $N_b$ increases with increasing $\delta$, see Ref.~\onlinecite{jpa}.  
The $N$-particle problem can then be studied for $N\le N_b(\delta)$ only. 
In fact, due to the repulsive interactions, the maximum number of 
bound electrons may be lowered even further.\cite{jpa}
For the infinite-mass confinement [case (i)], there is no constraint on
the number of particles held by the dot.
 
We now add electron-electron interactions to the $N$-particle 
problem.   The Coulomb interaction matrix 
elements are given in terms of the eigenspinors $\psi_a$,
\begin{equation}\label{genmat}
V_{aa'b'b} := \alpha \int \frac{\rd{\bm r} \rd{\bm r}'}{|{\bm r}-{\bm r}'|}
\left(\psi_a^\dagger \cdot \psi_b^{}\right)({\bm r})
\left(\psi_{a'}^\dagger \cdot \psi_{b'}^{}\right)({\bm r}'),
\end{equation}
with $V_{a'abb'}=V_{aa'b'b}$.
Due to total angular momentum conservation, only matrix
elements with $m_a+m_{a'}=m_b+m_{b'}$ do not vanish.
Interaction matrix elements with 
large momentum exchange $k=m_b-m_a$ ($k\in\mathbb{Z}$)
are numerically small,\cite{hausler} but all possible values of $k$
(for a chosen basis size) are taken into account below.
For the magnetic dot [case (ii)], the matrix elements \eqref{genmat} are most
conveniently evaluated by expanding $\psi_a$ in
conventional relativistic Landau level states.\cite{jpa}
For the infinite-mass confinement [case (i)], after inserting 
Eq.~\eqref{confmass} into Eq.~\eqref{genmat},
some algebra [cf.~also Appendix B of Ref.~\onlinecite{jpa}] yields 
\begin{widetext}
\begin{eqnarray}\label{genmathard}
V_{aa'b'b} &=& (4\pi)^2 \alpha A_{a} A_{a'}A_{b'} A_b \sum_{l=0}^\infty 
C_{k,l} \int_0^1 \rd r \ r^{-l} \left(J_{m_a}(E_a r) J_{m_b}(E_br) +
J_{m_a+1}(E_a r) J_{m_b+1}(E_b r)\right)\\ \nonumber 
&\times & \int_0^r \rd r' \ (r')^{l+1} \left(
J_{m_{a'}}(E_{a'} r') J_{m_{b'}}(E_{b'}r' ) 
+J_{m_{a'}+1}(E_{a'} r') J_{m_{b'}+1}(E_{b'}r' ) \right) .
\end{eqnarray}
\end{widetext}
The coefficient $C_{k,l}$ vanishes when $l+|k|$ is odd or when $l<|k|$.
For $k=l=0$, we have $C_{k,l}=1/2$. In all remaining cases, we
obtain
\[
C_{k,l} = \frac{(2l-1)!!}{2^{l+1} l!} \prod_{n=1}^{(l+|k|)/2} 
\frac{(n-1/2)(n-l-1)}{n(n-l-1/2)}.
\]
Equation \eqref{genmathard} is then evaluated by numerical integration
routines and yields the interaction matrix elements.

\subsection{Numerical approaches}

Next, we briefly describe three different numerical approaches 
to obtain the ground-state energy for a quantum dot containing $N$ 
Dirac fermions, namely HF simulations, the M\"uller density matrix formulation, 
and exact diagonalization (for $N=2$).  
In order to have a well-defined many-body problem, we follow 
Sucher\cite{sucher} and restrict ourselves to the
projected single-particle space, i.e., we assume an inert filled
Dirac sea.  Hence summations over $a=(m,n)$ 
will only include positive-energy single-particle solutions ($E_a>0$).
In the numerical calculations, the basis size (i.e., the number $K$ of
single-particle orbitals spanning the Hilbert space)
was always chosen sufficiently large to ensure convergence.  
Failure to converge indicates an instability of the method, as 
we will see in the case of the M\"uller functional for strong interactions.

First, the Hartree-Fock approach amounts to the self-consistent minimization
of the functional
\begin{equation}\label{hff}
E_{\rm HF} [\gamma] =
 \sum_a E_a \gamma_{aa} +\frac12
\sum_{aa'bb'} (V_{aa'b'b}-V_{aa'bb'}) \gamma_{a'b'}\gamma_{ab},
\end{equation}  
where the density matrix $\gamma$ obeys $\gamma^2=\gamma$ and 
${\rm tr}(\gamma)=N$.  In our case, $\gamma$ is a real symmetric matrix.
The numerical algorithm to obtain the HF ground state is standard
and can be found, for instance, in Ref.~\onlinecite{jpa}.
Second, the M\"uller density matrix formulation employs a 
different form for the exchange term, where one minimizes the functional\cite{muller,siedentop}
\begin{eqnarray}\label{mullerf}
E_{\rm M}[\gamma] &=& 
 \sum_a E_a \gamma_{aa} +\frac12
\sum_{aa'bb'} \Bigl( V_{aa'b'b} \gamma_{a'b'}^{}\gamma^{}_{ab}  \\ \nonumber
&-&  V_{aa'bb'} (\gamma^{1/2})_{a'b'} (\gamma^{1/2})_{ab} \Bigr),
\end{eqnarray}
where $\gamma$ is again a real symmetric matrix with ${\rm tr}(\gamma)=N$,
but now $\gamma^2\leq \gamma$.
A stable numerical approach to minimize $E_{\rm M}[\gamma]$ 
in Eq.~\eqref{mullerf}, the so-called
projected gradient algorithm, has been formulated and 
tested before.\cite{gritsenko,pernal} We 
have employed precisely the same method here.
Finally, the exact numerical diagonalization of the full many-body problem
is only possible for small particle numbers due to the exponential increase
in computational complexity with increasing $N$. 
We have therefore carried out exact
diagonalization calculations only for $N=2$ Dirac fermions, 
primarily to check the accuracy of the two computationally less expensive 
but approximate alternative approaches.
Details of the  exact diagonalization approach have been 
described in Ref.~\onlinecite{hausler}.

\section{Comparison of methods: $N=2$}
\label{sec3}

\begin{figure}[t!]
\begin{center}
\includegraphics[width=2.8in]{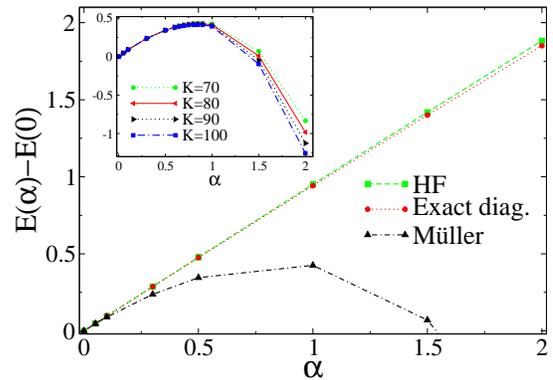}
\caption{\label{fig1} (color online) Interaction contribution to the
ground-state energy, $E(\alpha)-E(0)$ (in units of $\hbar v_F/R$),
vs fine structure constant $\alpha$ for the infinite-mass confined dot 
containing $N=2$ particles.  The main panel shows the results of the 
three different approaches, see main text.
The HF results are very close to the exact diagonalization results, while
the M\"uller functional gives a lower bound. 
Straight lines are a guide to the eye only.
 Inset: Same for 
M\"uller functional, with different basis size $K$. Note the
absence of convergence for large $\alpha$.  }
\end{center}
\end{figure}

In this section, we show  and compare the results of the
three approaches described in Sec.~\ref{sec2}
for $N=2$ Dirac fermions.  For the infinite-mass confinement case,
results for the ground-state density $E(\alpha)$ 
are shown in Fig.~\ref{fig1}.   Clearly, the numerically exact 
result obtained from exact diagonalization
is bracketed by the HF prediction from above and by the 
M\"uller result from below.  The HF approximation provides
very accurate estimates for $E(\alpha)$, while the M\"uller functional
is only reliable for very small $\alpha$.  In fact, the
application of the M\"uller functional to the free-space case
reveals an intrinsic divergence for 
strong interactions $\alpha> \alpha_c$, where the critical value is 
(see Ref.~\onlinecite{Walter2010}, correcting an earlier attempt\cite{Bouzouina}) 
\begin{equation}\label{ac}
\alpha_c = \frac{2}{y+1/y}\approx 0.756,\quad 
y=\frac{\Gamma^4(1/4)}{8\pi^2}.
\end{equation}
Although this critical value was derived for the case of vanishing
magnetic field, we anticipate that it applies also to the confined
geometry, with or without magnetic field, since it arises from the
fact that both the kinetic energy and the Coulomb singularity scale as
inverse length for short distances, i.e., a regular magnetic field is
clearly irrelevant for the singular behavior. (Such a result has
recently been established in a related situation.\cite{MaierSiedentop})
For $\alpha>\alpha_c$, the exchange part in the M\"uller functional
[Eq.~(\ref{mullerf})] provides a strong attraction which effectively
forces particles to form a droplet.  In our case, this singular
behavior implies that the ``ground-state'' energy drops to
$-\infty$. In numerical computations, this is reflected by the fact
that the energy becomes cutoff-dependent, going to $-\infty$ as
the basis size $K$ grows. This phenomenon is clearly visible in the
inset of Fig.~\ref{fig1}, but a precise comparison of the predicted
critical value for $\alpha_c$ [Eq.~\eqref{ac}] with numerics is
difficult. This singularity is an unphysical artefact of the M\"uller
density matrix approach and indicates that it is only useful for
$\alpha\ll 1$.  On the other hand, the HF approximation is very close
to the exact value even for $\alpha=2$.

\begin{figure}[t!]
\begin{center}
\includegraphics[width=2.8in]{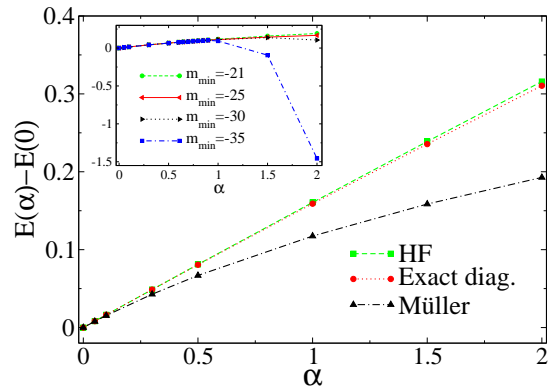}
\caption{\label{fig2} (color online) Same as Fig.~\ref{fig1} but for the
magnetically confined case with $\ell = R$.  
The basis size $K$ here corresponds to $2|m_{\rm min}|$.   
}
\end{center}
\end{figure}

A very similar picture emerges from the corresponding study
of the magnetically confined dot, see Fig.~\ref{fig2}.
In both cases and for all $\alpha\le 2$,
the interaction energy obtained under the HF approximation 
is less than  $1\%$ above the corresponding exact value.
In the remainder of the paper, we will then
study $N>2$ particles using the HF approach.  
We have compared the results of the M\"uller functional for $N>2$ 
to the corresponding HF results as well, and with increasing
$N$ they come closer.  Hence we expect that the relative accuracy
of the HF results (at the least) does not deteriorate for $N>2$.

\section{Hartree-Fock results for $N>2$ particles}\label{sec4}

In the previous section, we have established that HF calculations
are able to provide very accurate estimates for the ground-state energy of 
Dirac fermions in a circular quantum dot.  In this section, we 
describe the results of our HF calculations for up to $N=20$ particles.
For clarity, we focus on the infinite-mass confinement case, but
qualitatively similar results were also found for the magnetic confinement.

\begin{figure}[t!]
\begin{center}
\includegraphics[width=2.8in]{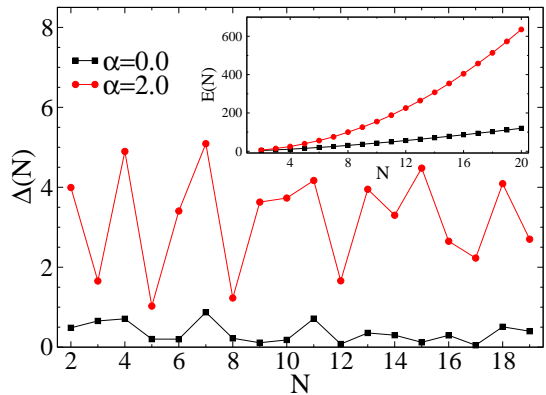}
\caption{\label{fig3} (color online)
HF results for the addition energy $\Delta(N)$ [Eq.~\eqref{add}]
vs particle number $N$ for a dot formed by infinite-mass confinement.  
Results are shown for $\alpha=2$ (red circles) and 
for $\alpha=0$ (black squares); straight lines are a guide to the eye
only.  Inset: HF results for the energy $E(N)$ vs $N$, for the same
interaction parameters. }
\vspace{1cm}
\end{center}
\end{figure}

Figure~\ref{fig3} shows HF results for the 
$N$-dependent addition energy,
\begin{equation}\label{add}
\Delta(N) := E(N+1) + E(N-1)-2 E(N),
\end{equation}
both for $\alpha=2$ and for the noninteracting case ($\alpha=0$). 
The HF ground-state energy $E(N)$ obtained from 
our self-consistent numerical calculation
is shown in the inset of Fig.~\ref{fig3}.
A peak in the addition energy for some $N$ implies a higher stability
of the $N$-particle dot. In analogy to atomic and nuclear
 physics, this $N$ is often
 referred to as ``magic number.''\cite{reimann}   
While already the noninteracting dot has some structure in the 
addition energy spectrum (due to the single-particle spectrum), 
e.g., the small peaks
at $N=7$ and $N=11$ visible in Fig.~\ref{fig3}, 
the interacting case is characterized
by more pronounced features. 
For $\alpha=2$, we observe clear peaks,
see Fig.~\ref{fig3}, corresponding
to the magic numbers $N=4,7, 11, 13, 15$ and $18$.  Although some of these
numbers coincide with the noninteracting ones, it is evident
that the addition energy spectrum is drastically changed by 
electron-electron interactions in such a finite-size system.  

\begin{figure}[t!]
\begin{center}
\includegraphics[width=2.8in]{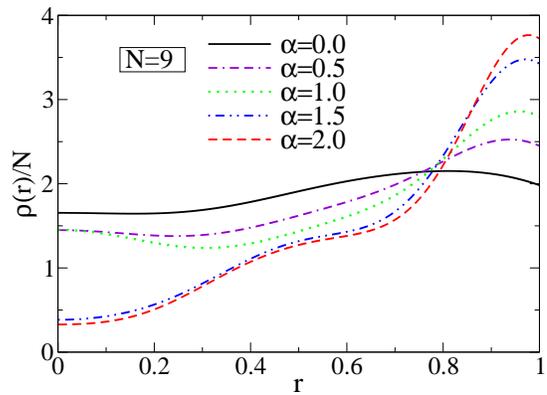}
\caption{\label{fig4} (color online)
HF results for the radial density profile $\rho(r)$ vs $r$ for $N=9$ particles 
with several $\alpha$.  }
\end{center}
\end{figure}

\begin{figure}[h]
\begin{center}
\includegraphics[width=2.8in]{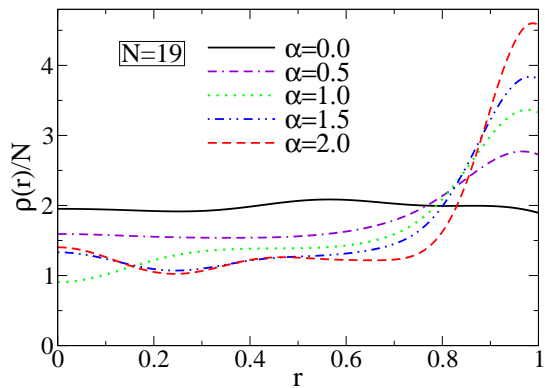}
\caption{\label{fig5} (color online)
Same as Fig.~\ref{fig4} but for $N=19$.
}
\end{center}
\end{figure}

The resulting ground-state density $\rho({\bm r})$ is rotationally 
invariant and can therefore be analyzed in terms of 
the angular-averaged density $\rho(r)$, which is normalized as
$\int_0^1 rdr \rho(r)=N$.
Figure \ref{fig4} shows HF results for the density
$\rho(r)$ for $N=9$ and several $\alpha$.  
In the noninteracting case ($\alpha=0$), 
the density profile is rather smooth, but with increasing $\alpha$ 
the particles are pushed towards the boundary and form a ring. 
When comparing the shoulder-like feature apparent in Fig.~\ref{fig4} (around
$r\approx 0.5$) to the corresponding correlation plot (see below), 
we find that no significant particle weight is contained in the
shoulder, i.e., with high probability all particles are close
to the boundary.
For $N=19$ particles, 
a richer structure emerges, see Fig.~\ref{fig5},
with three different spatial ``shells'' emerging for strong interactions. 
In particular, by integrating over the shown density curve, we find
that one particle is located near the origin, three particles are contained
in a second shell around $r\approx 0.45$, and the remaining 15 particles
are close to the boundary.

To obtain more detailed insight we next study
the density-density correlation function
\begin{equation}\label{gcor}
g({\bm r}, {\bm r}') = \langle \rho({\bm r} )\rho({\bm r}')\rangle, 
\end{equation}
where ${\bm r}'$ is kept fixed. Monitoring $g({\bm r},{\bm r}')$
as a function of ${\bm r}$,  the spatial arrangement
of the particles in the dot can be revealed. 

\vspace{2cm}

\begin{figure}[t!]
\begin{center}
\includegraphics[width=2.8in]{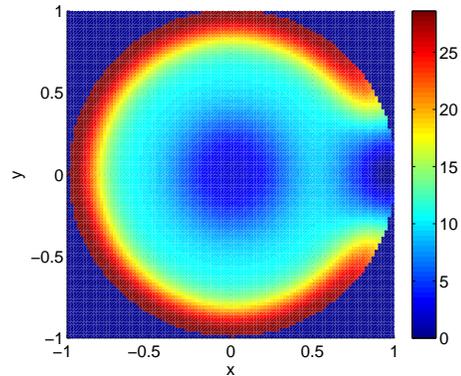}
\caption{\label{fig6} (color online)
Correlation plot $g({\bm r},{\bm r}')$ for $N=9$ particles and  $\alpha=2$, 
corresponding to Fig.~\ref{fig4}.  The position ${\bm r}'$ is fixed at
$(0.95,0)$, and the color scale indicates the correlation degree 
for different ${\bm r}$ within the quantum dot.  
}
\end{center}
\end{figure}

\begin{figure}[t]
\begin{center}
\includegraphics[width=2.8in]{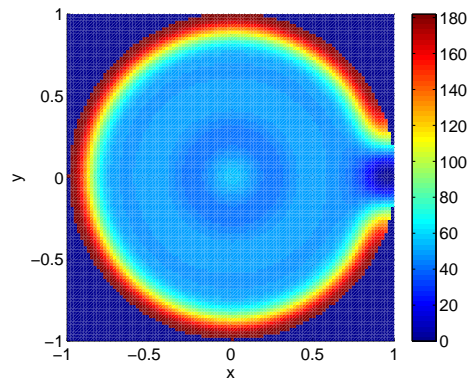}
\caption{\label{fig7} (color online)
Same as Fig.~\ref{fig6} but for $N=19$.  }
\end{center}
\end{figure}

2D correlation plots for $N=9$ and $N=19$ (with $\alpha=2$) are
shown in Fig.~\ref{fig6} and Fig.~\ref{fig7}, respectively.
In these plots, we keep ${\bm r}'=(0.95,0)$ fixed and show the
correlations as function of ${\bm r}=(x,y)$ within the dot. 
For $N=9$, Fig.~\ref{fig6} is consistent with all electrons being arranged
equidistantly on a ring close to the boundary.
The correlation plot in Fig.~\ref{fig7}
for $N=19$ particles also confirms the conclusions reached from the
analysis of the density plot in Fig.~\ref{fig5}.   
The outermost spatial shell (near the boundary) holds 15 particles,
a second ring contains 3 particles, and one particle 
is located at the center.
The combined analysis of density and correlation plots for all
particle numbers under study, $N\le 20$,
results in the shell filling sequence in Table \ref{table1}.

\vspace{1cm}

\vspace{1cm}

These observations provide a signature 
for the onset of Wigner molecule behavior, 
i.e., we have a finite-size system where Wigner crystallization sets in but
quantum fluctuations are still important.\cite{egger,uzi} 
In order to compare to  the deep Wigner crystallized limit, we 
now briefly discuss the classical limit (which here is defined by taking
the limit $\alpha\to \infty$), 
where the electrostatic energy dominates
completely and the kinetic energy can be neglected.
The repulsive interaction then tries to maximize the distance between
particles, leading to the formation of spatial shells.
 The shell filling sequence for a harmonically confined Wigner molecule
(of Schr\"odinger fermions) is well known,\cite{reimann,bedanov,egger}
and HF calculations have been able to capture the Wigner molecule formation.\cite{uzi}
For the 2D circular hard-wall confinement considered here,  however,
a different shell filling sequence follows by minimization of the 
classical electrostatic energy $E_c(N)$ 
with respect to all particle positions ${\bm r}_{i=1,\ldots,N}$ 
within the disk ($r_i\le R=1$),
\begin{equation}\label{ec1}
E_c(N) = \sum^N_{i<j} \frac{\alpha}{|{\bm r}_j-{\bm r}_i|}.
\end{equation}
A first possible configuration has all particles arranged
equidistantly on a unit circle, resulting in the classical energy
\begin{equation}
E_c^{(1)}(N) = \frac{N\alpha}{2} \times \left\{
\begin{array}{cc}  \sum_{k=1}^{(N-1)/2} \frac{1}{\sin(\pi k/N)}, & N
 \ {\rm odd},\\
\sum_{k=1}^{(N-2)/2} \frac{1}{\sin(\pi k/N)} + \frac12 ,& N \ {\rm even}.
\end{array} \right.
\end{equation}
If instead one particle resides at the origin plus $N-1$ particles
on the outer ring as above, the energy of this second configuration is
\begin{equation}
E_c^{(2)}(N) = E_c^{(1)}(N-1) +(N-1)\alpha.
\end{equation}

\vspace{1cm}
\begin{table}[t!]
 \centering
 \begin{tabular}{|c||c|c|c|| c|c|c|}\hline
$N$ & $N_1$ & $N_2$ & $N_3$ & $N^{\rm cl}_1$ & $N^{\rm cl}_2$ & 
$N^{\rm cl}_3$  \cr\hline\hline
2  & 2 & - & - & 2 & - & - \cr\hline
3  & 3 & - & - & 3 &- &- \cr\hline
4  & 4 & - & - & 4 &- &- \cr\hline
5  & 4 & - & - & 5 &- &- \cr\hline
6  & 6 & - & - & 6 &- & - \cr\hline
7  & 7 & - & - & 7 &- & - \cr\hline
8  & 8 & - & - & 8 &- &- \cr\hline
9  & 9 & - & - & 9 &- &-  \cr\hline
10  & 10 & - & - & 10 &- &- \cr\hline
11  & 1 & 10 & - & 11  &- &- \cr\hline
12  & 1 & 11 & - & 1 & 11 & -\cr\hline
13  & 1 & 12 & - & 1 & 12 & - \cr\hline
14  & 1 & 13 & - & 1 & 13 & - \cr\hline
15  & 2 & 13 & - & 1 &14 &- \cr\hline
16  & 3 & 13 & - & 1 &15 &-  \cr\hline
17  & 1 & 3 & 13 & 2 & 15 & - \cr\hline
18  & 1 & 3 & 14 & 2 & 16 & - \cr\hline
19  & 1 & 3 & 15 & 3 & 16 & -\cr\hline
20  & 1 & 3 & 16 & 3& 17& - \cr\hline
 \end{tabular}
 \caption{Shell filling sequence for a  2D
interacting Dirac fermion dot with circular hard-wall confinement.  
$N_{i}^{\rm cl}$ denotes
the number of particles in the $i$th spatial shell obtained
from the minimization of the classical electrostatic energy [Eq.~(\ref{ec1})].
$N_i$ is the corresponding HF quantity for $\alpha=2$,  
see main text.}
\vspace{1cm}
\label{table1}
\end{table}

For $N\leq 16$, numerical minimization of Eq.~(\ref{ec1}) shows
that these two configurations always yield the lowest-energy
solutions. In particular,  $E_c^{(1)}<E_c^{(2)}$ for $N<12$, 
see Table \ref{table1}. 
For $16<N\leq 20$, an additional inner ring is formed 
containing $N_1^{\rm cl}>1$ particles, surrounded by the outer 
ring containing $N_2^{\rm cl}=N-N_1^{\rm cl}$ particles.  
For all $N\le 20$, the classical lowest-energy solution thus has at most
two spatial shells, but configurations with three shells as observed
in the quantum calculation are energetically quite close.
The agreement between the shell filling sequence observed for $\alpha=2$
and in the classical limit is not perfect but indicates that we
are already rather close to the classical limit for $\alpha=2$
and have a Wigner molecule, despite of the theoretically predicted absence of
Wigner crystallization in bulk graphene.\cite{wigner}
Even for $\alpha=1$, the above density plots suggest that incipient
Wigner molecule behavior can be observed.  (Of course, this is a smooth
crossover and not a phase transition.)
However, the fact that there are still substantial quantum fluctuations
for $\alpha=2$ is also clear from the addition spectrum in Fig.~\ref{fig3}.
In the deep classical limit, there is much less pronounced structure
in the addition energy spectrum.

\begin{figure}[t!]
\begin{center}
\includegraphics[width=2.8in]{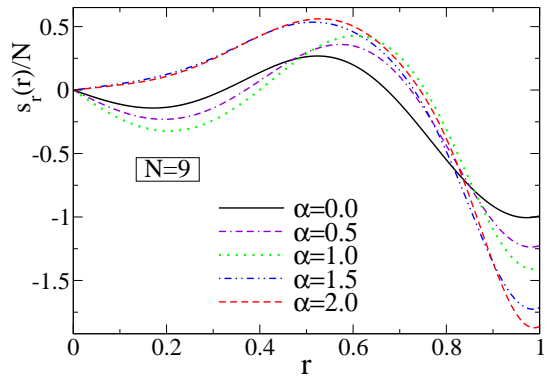}
\caption{\label{fig8} (color online) HF results for the
spin density $s_r$ in radial direction vs $r$ for $N=9$ and various $\alpha$.
}
\end{center}
\end{figure}

Finally, we point out that there is also interesting spin texture in
such a quantum dot.  The Pauli matrices in Eq.~(\ref{d0}) are 
directly connected to the electronic spin density in 
a topological insulator surface
via the relation\cite{hasan,scatt} 
\begin{equation}\label{sdens}
{\bm s}({\bm r})= (s_x,s_y)^T = \frac{\hbar}{2} \langle \hat e_z \times 
{\bm \sigma}\rangle. 
\end{equation}
For the case of graphene, the Pauli matrices refer to the sublattice
degree of freedom, which is not easily accessible to experiments.
The spin density (\ref{sdens}) points within the 2D plane 
and is always isotropic, i.e., independent of the angular variable.
We find that only the radial component $s_r(r) :={\bm s}\cdot \hat e_r$ (with
$\hat e_r={\bm r}/r$) does not vanish.  The resulting nontrivial
spin texture is shown in Fig.~\ref{fig8} for $N=9$ and several $\alpha$.

\section{Discussion}\label{sec5}

In this paper, we have discussed interaction effects in circular 2D
quantum dots where the particles are massless Dirac fermions.  Physical
realizations of the studied model are given by graphene and the surface
state of a topological insulator.
For the case of two particles, we have compared three different methods
to establish that Hartree-Fock calculations provide highly accurate
results for physically relevant interaction strengths.  
An alternative method based on the M\"uller density matrix functional was
also studied, but since M\"uller's \textit{Ansatz} 
for the two-particle density
respects the right normalization condition but sacrifices its
positivity, it suffers from an unphysical divergence for sufficiently
strong interactions. An improvement would have to take this drawback
into account, while not dropping the sum rule for the density and the
convexity of the functional.

The case of $N\le 20$ particles has then been studied using Hartree-Fock
simulations.  The resulting addition spectrum of the quantum dot 
reveals pronounced magic numbers that cannot be explained by a noninteracting
picture.  Moreover, the density profiles and the density-density
correlation functions show that we are rather close to the  classical
limit already for experimentally relevant interaction 
parameters ($\alpha\approx 1$ to $2$).  The formation of spatial shells
is a clear signature of a Wigner molecule, and we therefore predict
that in such a finite-size system the usual argument\cite{wigner}
for absence of Wigner crystallization of Dirac fermions can 
be effectively circumvented.

\acknowledgments 

We acknowledge useful discussions with A. De Martino, W. H\"ausler,
 and E. Stockmeyer.
This work was supported by the Sonderforschungsbereich TR 12 of the DFG.

\end{document}